\documentclass{iopconfser}
\usepackage[utf8]{inputenc} 
\usepackage[T1]{fontenc}    
\usepackage{hyperref}       
\usepackage{url}            
\usepackage{booktabs}       
\usepackage{amsfonts}       
\usepackage{nicefrac}       
\usepackage{microtype}      
\usepackage{xcolor}         

\usepackage{amsmath}         
\usepackage{braket}
\usepackage{algorithm}
\usepackage{algpseudocode}
\usepackage{graphicx,wrapfig,lipsum}
\usepackage{amssymb}
\usepackage{wrapfig}
\usepackage{enumerate}

\begin{document}

\title{Are queries and keys always relevant? A case study on Transformer wave functions}

\author{Riccardo Rende$^{1,*}$, Luciano Loris Viteritti$^{2,*}$}

\affil{$^1$International School for Advanced Studies, Trieste, Italy}
\affil{$^2$University of Trieste, Trieste, Italy}

\email{rrende@sissa.it, lucianoloris.viteritti@phd.units.it}

\begin{abstract}
The dot product attention mechanism, originally designed for natural language processing tasks, is a cornerstone of modern Transformers. It adeptly captures semantic relationships between word pairs in sentences by computing a similarity overlap between queries and keys. In this work, we explore the suitability of Transformers, focusing on their attention mechanisms, in the specific domain of the parametrization of variational wave functions to approximate ground states of quantum many-body spin Hamiltonians. Specifically, we perform numerical simulations on the two-dimensional $J_1$-$J_2$ Heisenberg model, a common benchmark in the field of quantum many-body systems on lattice. By comparing the performance of standard attention mechanisms with a simplified version that excludes queries and keys, relying solely on positions, we achieve competitive results while reducing computational cost and parameter usage. Furthermore, through the analysis of the attention maps generated by standard attention mechanisms, we show that the attention weights become effectively input-independent at the end of the optimization. We support the numerical results with analytical calculations, providing physical insights of why queries and keys should be, in principle, omitted from the attention mechanism when studying large systems.
\end{abstract}

\section{Introduction}
\let\thefootnote\relax\footnotetext{*Equal contribution.}
Transformers~\cite{vaswani2017} have emerged as one of the most powerful deep learning tools in recent years. They are task-agnostic neural networks that are pre-trained to build context-sensitive representations of words in input sentences~\cite{devlin2019bert,radford2018improving,radford2019language}. The success of Transformers lies in their remarkable flexibility: with minimal modifications, they excel in addressing diverse problem domains, often outperforming specialized approaches~\cite{jumper2021highly,chatgpt,dosovitskiy2021an}. This is a consequence of their versatile foundational components, namely the dot product self-attention mechanism, Multilayer Perceptron (MLP), Layer Normalization, and skip connections. While elements like the MLP, Layer Normalization, and skip connections are task-agnostic and offer broad applicability, the functional form of the dot product attention mechanism was originally tailored for natural language processing (NLP) tasks. In this context, a sentence is processed by initially associating each word with a vector through a lookup table. These vectors form a sequence which is processed by the self-attention mechanism~\cite{vaswani2017}, designed to generate, for each input, an output vector as a weighted sum of all other inputs. Crucially, the coefficients in this sum involve learnable parameters that are optimized to capture the semantic relationships between pairs of words within the sentence. The remarkable generalization properties of Transformers in NLP tasks have been associated with the use of attention weights that depend on the input values, thereby capturing powerful inductive biases related to the semantics in natural languages~\cite{clark2019does,bertology2021}. One wonders if the dot product attention mechanism provides an inductive bias which is the most appropriate in \emph{any} data domain. For example, Ref.~\cite{bhattacharya2020single} in the context of protein contact prediction and Ref.~\cite{jelassi2022} in computer vision tasks suggest that input-independent attention weights achieve competitive performance compared to the standard approach.  In this paper, we delve into this aspect by exploring the application of the Transformer architecture as a Neural-Network Quantum State (NQS) for approximating the ground state of quantum many-body spin Hamiltonians on lattice~\cite{carleo2017}. The Transformer architecture has already been employed in this context, achieving highly accurate results across different systems~\cite{melko2024language, czischek2023, rende2023, viteritti2023, luo2022, diluo2023, viteritti20231d,vonglehn2023}. While many of these works adopt the standard attention mechanism~\cite{czischek2023, luo2022, diluo2023}, Ref.~\cite{rende2023} employs a simplified version, omitting queries and keys, still reaching \textit{state-of-the-art} accuracy on one of the most popular benchmark problems in frustrated magnetism. Therefore, the question of whether queries and keys provide a suitable inductive bias for general applications persists. In this work, we tackle this question by systematically investigating the performance of different attention mechanisms within Transformer wave functions. In the following, we summarize our main findings:

\begin{enumerate}[(i)]
    \item In Transformer wave functions, the standard dot-product attention mechanism used in NLP does not improve the performance of a simpler mechanism in which the attention weights are input-independent. 
    \item By analyzing the attention maps produced by architectures including queries and keys, we find that the optimization process makes them efficaciously input-independent.
    \item Based on analytical computations, we provide insights into why conventional attention mechanisms are expected to converge towards input-independent solutions when applied to systems which are sufficiently large to be split in independent subsystems. 
\end{enumerate}
Interestingly, the result of point (iii) can be extended to other domains, such as NLP or computer vision, in cases where tasks can be solved by exploiting correlations over shorter lengths compared to the entire input sequence, thereby partitioning the input into effectively uncorrelated parts.

\section{Background}
\subsection{The quantum-many body problem}
The physical properties of an interacting quantum-many body system described by a Hamiltonian $\hat{H}$ are determined by solving the time-independent Scr\"{o}dinger equation $\hat{H}\ket{\Psi_n}=E_n\ket{\Psi_n}$, where $\ket{\Psi_n}$ and $E_n$ are eigenstates and eigenvalues of $\hat{H}$, respectively. In principle, fixing a basis in the Hilbert space, we can numerically obtain the spectrum of $\hat{H}$ by storing all its matrix elements and using standard computational routines to diagonalize it. However, a critical challenge arises due to the exponential growth in the size of this matrix with respect to the number of particles in the system, rendering this approach feasible only for small systems~\cite{sandvik2010}. Typically, the focus lies in the low-energy properties of the Hamiltonian, particularly in its ground state \(\ket{\Psi_0}\). To obtain approximations of the ground state for systems where exact diagonalization is not feasible, many methods have been developed over the years. Here, we focus on variational approaches, where a variational state \(\ket{\Psi_{\theta}},\) depending on a set of $N_p$ parameters \(\theta\), is optimized to minimize the variational energy \(E_{\theta} = \braket{\Psi_{\theta}|\hat{H}|\Psi_{\theta}}/\braket{\Psi_{\theta}|\Psi_{\theta}}\). According to the Variational Principle~\cite{sakurai2020}, the energy $E_{\theta}$ associated to any generic state \(\ket{\Psi_{\theta}}\) is always bigger than the ground state energy \(E_{\theta} \ge E_0\). Moreover, provided that the ground state is unique, we have that \(E_{\theta} = E_0\) if and only if \(\ket{\Psi_{\theta}} = \ket{\Psi_0}\). To be concrete, we consider systems of \(N\) spin-\(1/2\) arranged on regular lattices. In this case, the variational state can be expanded as \(\ket{\Psi_{\theta}} = \sum_{\{\sigma\}} \Psi_{\theta}(\sigma) \ket{\sigma}\), where \(\{\ket{\sigma} = \ket{\sigma_1^z, \sigma_2^z, \dots, \sigma_N^z}\}\) with \(\sigma^z_i = \pm 1\) is the computational basis. The many-body wave function \(\Psi_{\theta}(\sigma) = \braket{\sigma | \Psi_{\theta}}\) is a compact representation of the quantum state, which maps configurations of the basis set \(\ket{\sigma}\) to complex numbers using a relatively small number of parameters $N_p$ compared to the exponential size of the full Hilbert space (\(N_p \ll 2^N\)). 

\subsection{Variational Monte Carlo Framework} \label{subsec:VMC}
The Variational Monte Carlo (VMC) is a general framework used to construct an approximation of the ground-state \(\ket{\Psi_{0}}\) of a quantum many-body Hamiltonian \(\hat{H}\)~\cite{becca2017}. This is achieved by minimizing the variational energy \(E_{\theta}\), associated with a trial variational state \(\ket{\Psi_{\theta}}\), through a gradient-based iterative procedure which employs stochastic estimations of the relevant quantities (see Algorithm~\ref{algorithm}). 
\begin{algorithm}
  \begin{algorithmic}[1]
       \State \textbf{Require:} Define a variational state $\Psi_{\theta}(\sigma)$
      \State \textbf{Require:} Initialize randomly the variational parameters $\theta$
      \For{$t = 1, N_{opt}$} \label{forloop}
        \State samples $\{\sigma_i\}_{i=1}^M \sim |\Psi_{\theta}(\sigma)|^2$ via MCMC
        \State Stochastic estimation of the gradient of the energy : $F_{\gamma} =- \partial_{\gamma} E_{\theta}$ with $\gamma=1, \dots, N_p$ 
        \State Stochastic estimation of the Quantum Geometric Tensor : $S_{\gamma, \beta}$ with $\gamma, \beta =1, \dots, N_p$
        \State Update of the parameters with Stochastic Reconfiguration: $\delta \theta_{\gamma} = \tau \sum_{\beta} S^{-1}_{\gamma, \beta} F_{\beta}$ 
        \State New parameters : $\theta \gets \theta + \delta \theta$
      \EndFor
  \end{algorithmic}
  \caption{Variational Monte Carlo} \label{algorithm}
\end{algorithm}
The key object of the algorithm is the gradient of the energy with respect to the variational parameters (see step 5 in Algorithm~\ref{algorithm}), which can be expressed as a correlation function~\cite{becca2017,carleo2017,rende2023}:
\begin{equation}\label{eq:grad_E}
    F_{\gamma} = -\frac{\partial E_{\theta}}{\partial \theta_{\gamma}} = - 2 \Re\left[ \braket{ (\hat{H} - \braket{\hat{H}})(\hat{O}_{\gamma} - \braket{\hat{O}_{\gamma}})} \right] \ ,
\end{equation}
where $\gamma=1, \dots, N_p$ and $\hat{O}_{\gamma}$ are diagonal operators defined as ${O_{\gamma} (\sigma) = {\partial}\text{Log}[\Psi_{{\theta}}(\sigma)]/{\partial \theta_{\gamma}}}$. The latter log-derivative can be efficiently computed for NQS architectures using automatic differentiation~\cite{jax2018github}. The expectation values \(\braket{\dots}\) are are stochastically estimated using Markov Chain Monte Carlo (see Appendix~\ref{app:monte_carlo}) by sampling $M$ configurations according to the amplitudes $|\Psi_{\theta}(\sigma)|^2$ (details can be found in Appendix~\ref{app:metropolis}). The parameters are updated according to the Stochastic Reconfiguration (SR) method~\cite{sorella1998,sorella2005} (see step 7 in Algorithm~\ref{algorithm}), which is formally equivalent to Natural Gradient~\cite{Amari1998,Amari2018}. The SR approach takes into account the geometric properties of the energy landscape through the Quantum Geometric Tensor $S$, a $P \times P$ matrix which generalizes the Fisher information metric~\cite{park2020}:
\begin{equation}\label{eq:S_matrix}
    S_{\gamma, \beta} = \Re\left[\braket{(\hat{O}_{\gamma} - \braket{\hat{O}_{\gamma}})^{\dagger}(\hat{O}_{\beta} - \braket{\hat{O}_{\beta}})} \right]  \ .
\end{equation}
Recent studies have demonstrated the effectiveness of SR in optimizing NQS with a large number of parameters~\cite{chen2023,rende2023,viteritti2023}. It is important to stress that in VMC the data, i.e., spin configurations, are generated ``on the fly'' during the optimization process by sampling from \(|\Psi_{\theta}(\sigma)|^2\). This is different from conventional machine learning scenarios where a fixed training set is provided.

\subsection{Vision Transformer wave function} \label{subsec:ViT}
In 2017, Carleo and Troyer~\cite{carleo2017} proposed using neural networks to parametrize the variational quantum state amplitudes $\Psi_{\theta}(\sigma) \in \mathbb{C}$. Neural-Network Quantum States have demonstrated remarkable representational power in challenging problems~\cite{glasser2018,lange2024} and reached state-of-the-art results in describing the ground state properties of two-dimensional frustrated magnets~\cite{nomuraimada2021,rende2023,viteritti2023,chen2023,roth2023}, bosonic~\cite{denis2024} and fermionic~\cite{robledomoreno2022,kim2023,pfau2020,nys2024} models. In this work, we focus on a particular NQS based on the Vision Transformer (ViT) architecture, introduced in Ref.~\cite{viteritti2023}. First, following what is done for images~\cite{dosovitskiy2021an}, each $L\times L$ input spin configuration $\sigma$ is split into patches of size $b \times b$, which are linearly embedded in a $d$-dimensional space, thus producing a sequence of $n=L^2/b^2$ vectors $(\boldsymbol{x}_1, \dots, \boldsymbol{x}_n)$, with $\boldsymbol{x}_i \in \mathbb{R}^d$. This sequence is processed by a deep ViT with real-valued parameters that produce an output sequence of vectors $(\boldsymbol{y}_1, \dots, \boldsymbol{y}_n)$, with $\boldsymbol{y}_i \in \mathbb{R}^d$.  The ViT architecture is constituted by $n_l$ encoder blocks, each of them including Multi-Head attention with $h$ heads, two-layer MLP with GeLU activation, skip connections and Pre-Layer Normalization~\cite{xiong2020}. Then, a $d$-dimensional hidden representation is obtained as $\boldsymbol{z}=\sum_{i=1}^{n} \boldsymbol{y}_i \in \mathbb{R}^d$. Only at the end, the latter is mapped to a complex number representing the logarithm of the amplitude. This final mapping is performed by an output layer parametrized as a shallow network, namely $\text{Log}[\Psi_{\theta}(\sigma)] = \sum_{\beta=1}^d  g( b_{\beta} + \boldsymbol{w}_{\beta} \cdot \boldsymbol{z} )$, with non-linearity $g(\cdot)=\text{logcosh}(\cdot)$ and complex-valued trainable parameters $\{b_{\beta}, \boldsymbol{w}_{\beta}\}_{\beta=1}^{d}$. For more details about the architecture see Ref.~\cite{viteritti2023}. 

\section{Methods}
\subsection{Relative positional attention mechanisms} \label{subsec:attn_mechanisms}
The success of the Transformer architecture is commonly attributed to the attention mechanism~\cite{vaswani2017}. The basic idea of the attention mechanism is to process an input sequence of $n$ vectors $(\boldsymbol{x}_1, \dots, \boldsymbol{x}_n)$, with $\boldsymbol{x}_i \in \mathbb{R}^d$, producing a new sequence $(\boldsymbol{A}_1, \dots, \boldsymbol{A}_n)$, with $\boldsymbol{A}_i \in \mathbb{R}^d$. The goal of this transformation is to construct context-aware output vectors by combining all input vectors~\cite{vaswani2017}:
\begin{equation}
    \boldsymbol{A}_i = \sum_{j=1}^n \alpha_{ij}(\boldsymbol{x}_i, \boldsymbol{x}_j) V \boldsymbol{x}_j \ .
\end{equation}
The attention weights $\alpha_{ij}(\boldsymbol{x}_i, \boldsymbol{x}_j)$ form a $n\times n$ matrix, where $n$ is the number of patches, which measure the relative importance of the $j$-$th$ input when computing the new representation of the $i$-$th$ input.
\begin{figure}[t]
\center
 \includegraphics[width=0.7\columnwidth]{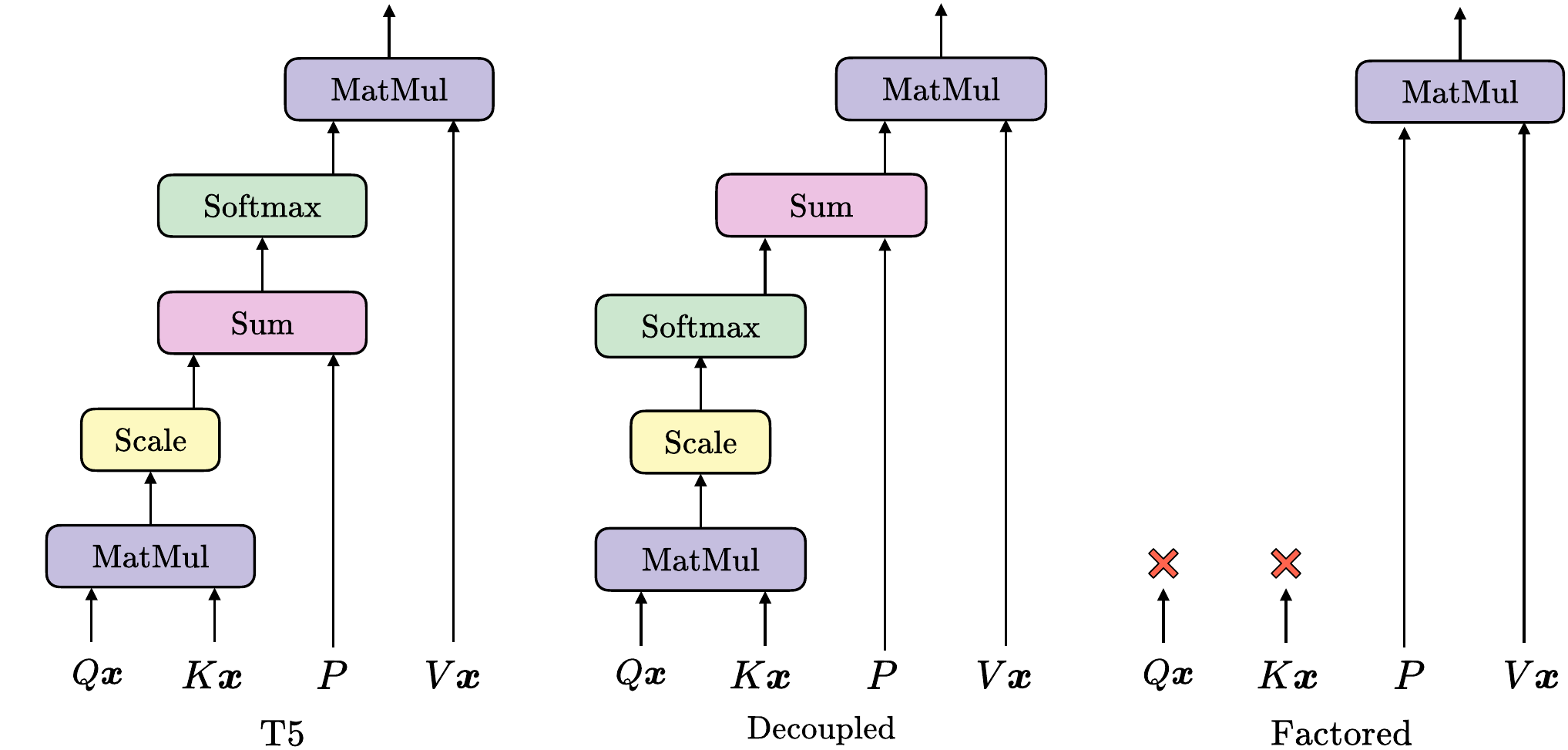}
 \caption{Schematic representation of the attention mechanisms employed in this work: T5~\cite{raffel2023exploring} (left panel), Decoupled~\cite{dai2021coatnet} (central panel) and Factored~\cite{bhattacharya2020single,rende2024mapping} (right panel) attention. In each of them, relative positional encoding is used. 
 The matrices $Q$, $K$, $V$ and $P$ are referred to queries, keys, values and positional encoding matrix, respectively.  Refer to Eqs.~\eqref{eq:t5},\eqref{eq:decoupled} and \eqref{eq:factored} in the main text for the analytical expressions.} \label{fig:attentions}
\end{figure}
During the years, several works proposed different parametrizations of the attention weights~\cite{shaw2018selfattention,wennberg2021case,ke2021rethinking}. Here, we consider three different mechanisms, all based on relative positional encoding~\cite{shaw2018selfattention}, as appropriate for the objective of this work.
\begin{enumerate}
\item \textit{T5 attention}, introduced in Ref.~\cite{raffel2023exploring}, is one of the most popular attention mechanisms:
\begin{equation}
\label{eq:t5}
    \alpha^{T5}_{ij}(\boldsymbol{x}_i, \boldsymbol{x}_j) = \frac{ \exp \left( \frac{ \boldsymbol{x}^T_i Q^T K \boldsymbol{x}_j}{\sqrt{d}} + p_{i-j} \right) }{ \sum_{k=1}^n \exp \left( \frac{\boldsymbol{x}^T_i Q^T K\boldsymbol{x}_k}{\sqrt{d}} + p_{i-k} \right) } \ .
\end{equation}
\item \textit{Decoupled attention}, introduced in Ref.~\cite{dai2021coatnet}:
\begin{equation}
\label{eq:decoupled}
    \alpha^{D}_{ij}(\boldsymbol{x}_i, \boldsymbol{x}_j) = \frac{\exp \left( \frac{\boldsymbol{x}^T_i Q^T K \boldsymbol{x}_j}{\sqrt{d}} \right) }{ \sum_{k=1}^n \exp \left( \frac{ \boldsymbol{x}^T_i Q^T K \boldsymbol{x}_k}{\sqrt{d}} \right) } + p_{i-j} \ .
\end{equation}
\item \textit{Factored attention}, introduced in Refs.~\cite{jelassi2022, bhattacharya2020single,rende2024mapping}:
\begin{equation}
\label{eq:factored}
        \alpha^{F}_{ij}(\boldsymbol{x}_i,\boldsymbol{x}_j) = p_{i-j} \ .
\end{equation}
\end{enumerate}
The vectors $Q\boldsymbol{x}_i$, $K\boldsymbol{x}_i$ and $V\boldsymbol{x}_i$ are called queries, keys and values, respectively. The matrices $Q$, $K$ and $V$, along with the positional encoding $P$, are trainable parameters.
When using relative positional encoding, the matrix $P$ is a \textit{circulant matrix} with dimensions $n \times n$ which is constructed by different circular shifts of a vector of parameters in different rows. This results in only $n$ independent trainable parameters, denoted by $p_{i-j}$. In Fig.~\ref{fig:attentions}, we show a schematic representation of these three different attention mechanisms. The Factored version has a reduced number of parameters, being the attention weights input independent. Regarding the computational cost for the calculation of each attention weight, we have $O(1)$ complexity in the Factored case and $O(n d^2) + O(n^2 d)$ in the other two cases. Decoupled attention, as represented by Eq.~\eqref{eq:decoupled}, is the simplest extension of the Factored version in Eq.~\eqref{eq:factored}, where the attention weights now factor in the input dependence: setting \(Q = K = 0\) allows recovering the Factored attention, albeit with a constant shift. Instead, in T5 attention [see Eq.~\eqref{eq:t5}] all the attention weights are constrained to be positive due to the global softmax activation.

\section{Results}
\begin{figure}[t]
\center
 \includegraphics[width=\columnwidth]{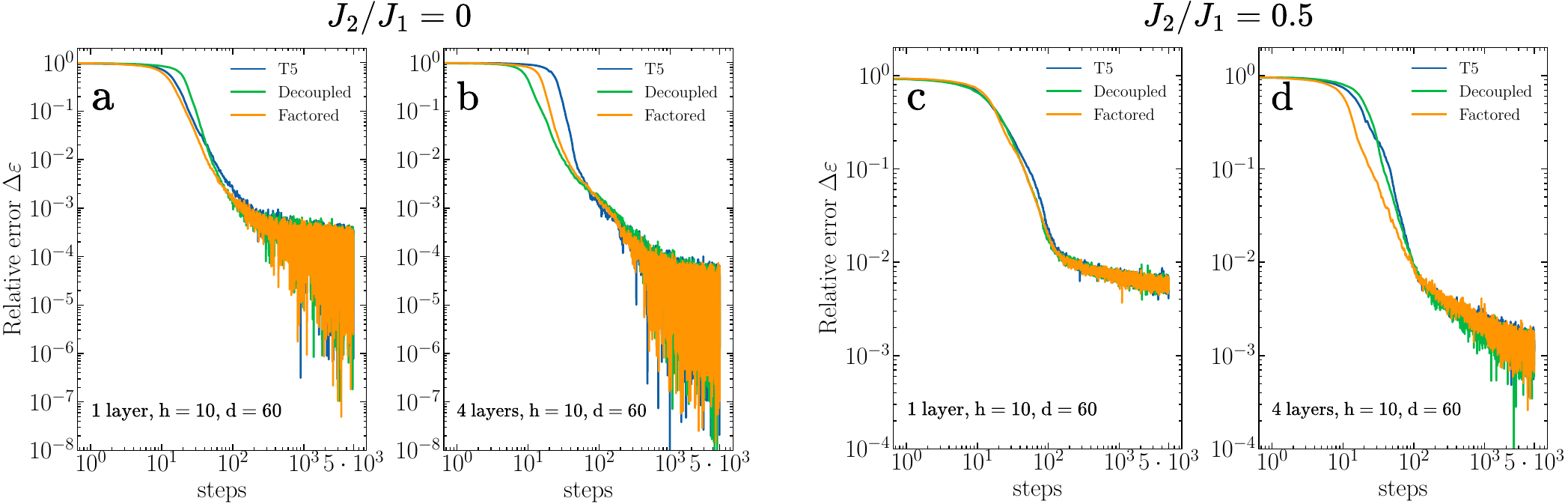}
 \caption{Relative error $\Delta \varepsilon = |(E_0 - E_{\text{ViT}})/E_0|$ during the optimization of the ViT wave function on the $J_1$-$J_2$ Heisenberg model at $J_2/J_1=0$ (left panel) and at $J_2/J_1 = 0.5$ (right) on a ${6 \times 6}$ lattice with periodic boundary conditions. The exact energies $E_{0}$ are computed with exact-diagonalization approaches. The architectures used for the simulations have $h=10$ heads, embedding dimension $d=60$, linear patch size $b=2$, $n_l=1$ layer in panels (a),(c), and $n_l=4$ layers in panels (b),(d). All networks are trained with the same optimization protocol, using SR (see section~\ref{subsec:VMC}) for $5\times 10^3$ optimization steps with $M=6\times10^3$ samples for the stochastic estimates. Each optimization step corresponds to one iteration in the for loop in Algorithm~\ref{algorithm}. A cosine decay learning rate scheduler is applied, starting with an initial value of $\tau=0.03$. The optimization curves are consistent across multiple runs with different random initialization of the parameters.} \label{fig:opt} 
\end{figure}
\subsection{Numerical experiments}\label{subsec:numerical_experiments}
We consider the two-dimensional $J_1$-$J_2$ Heisenberg model on a $L \times L$ square lattice, described by the following Hamiltonian:
\begin{equation}\label{eq:J1J2_ham}
  \hat{H} = J_1\sum_{\langle i,j \rangle} \hat{\boldsymbol{S}}_{i}\cdot\hat{\boldsymbol{S}}_{j} + J_2\sum_{\langle\langle i,j \rangle\rangle} \hat{\boldsymbol{S}}_{i}\cdot\hat{\boldsymbol{S}}_{j} \ ,
\end{equation}
where $\hat{\boldsymbol{S}}_{i}=(S_i^x,S_i^y,S_i^z)$ and $J_1,J_2 \ge 0$ are antiferromagnetic couplings for nearest- and next-nearest neighbors, respectively. 
The ground state of this model exhibits magnetic order in the two distinct limits $J_2/J_1 \ll 1$ and $J_2/J_1 \gg 1$. Specifically, when $J_2=0$ ($J_1=0$) the model reduces to the unfrustrated Heisenberg model, characterized by long-range Néel (columnar) magnetic order~\cite{calandra1998, sandvik1997}. In the intermediate region, particularly around \(J_2/J_1 \approx 0.5\), the system becomes highly frustrated, giving rise to exotic phases of matter~\cite{savary2017}.
The determination of the precise nature of the ground state in the frustrated region remains challenging and subject to debate~\cite{BeccaGutz2013, gong2014, nomuraimada2021}.

We employ a ViT wave function (see section~\ref{subsec:ViT}) to approximate, in the VMC framework (see section~\ref{subsec:VMC}), the ground state of this model on a \(L \times L\) lattice with periodic boundary conditions.
In order to assess the efficacy of the three distinct attention mechanisms introduced in section~\ref{subsec:attn_mechanisms}, we perform simulations on a $6 \times 6$ cluster utilizing ViT architectures with identical hyperparameters (embedding dimension $d$, number of heads $h$, number of layers $n_l$, and linear patch size $b$), modifying only the attention mechanism, namely T5 [see Eq.~\eqref{eq:t5}], Decoupled [see Eq.~\eqref{eq:decoupled}], and Factored [see Eq.~\eqref{eq:factored}]. In Fig.~\ref{fig:opt}, we report the optimization curves of the relative error of the variational energy with respect to the exact ground-state energy as a function of the optimization steps. On the left, we present the results for the unfrustrated case $(J_2/J_1=0)$ using ViT architectures with one [panel (a)] and four [panel (b)] layers. Instead, on the right, we report the results in the frustrated regime ($J_2/J_1=0.5$), again using one [panel (c)] and four [panel (d)] layers architectures. We emphasize that, although it is possible to enhance the performance of the variational state by employing larger architectures, such as increasing the number of layers, considering larger embedding dimensions or augmenting the number of heads~\cite{rende2023, viteritti2023, viteritti20231d}, the use of T5 or Decoupled attention mechanisms with input-dependent attention weights, and the subsequent increase of computational complexity and parameter count via the matrices \(Q\) and \(K\), does not produce improved results compared to Factored attention with input-independent attention weights. Notably, not only are the final accuracies practically identical, but also the learning dynamics exhibit similar behavior.

\begin{figure}
\centering
\includegraphics[width=\textwidth]{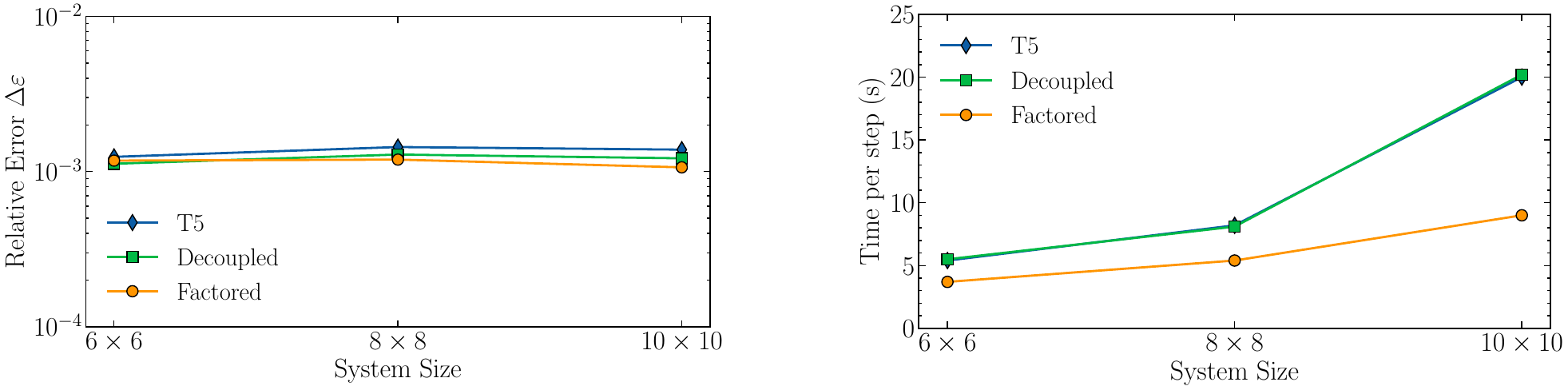}
\caption{\label{fig:size_scaling_L} \textbf{Left panel:} Relative error $\Delta \varepsilon = |(E_0 - E_{\text{ViT}})/E_0|$ at $J_2/J_1=0.5$ as a function of the system size for ViT architectures with the three different attention mechanisms, namely Factored (orange circles), Decoupled (green squares) and T5 (blue diamonds). The reference ground state energies are taken from exact diagonalization for $L=6$ $(-0.503810)$~\cite{schulz1996} and from variance extrapolation for $L=8$ ($-0.49906$)~\cite{BeccaGutz2013} and $L=10$ ($-0.497715$)~\cite{chen2023}. \textbf{Right panel:} Time per optimization step (in seconds) measured on a single GPU A100 for the three attention mechanisms as a function of the system size. For all simulations a ViT architecture with hyperparameters $d=10$, $h=10$, $b=2$ and $n_l=4$ is considered. The model is optimized using the SR optimization method for $5 \times 10^3$ steps, employing $M=6\times10^3$ training samples (see section~\ref{subsec:VMC}). A cosine decay learning rate scheduler is applied, starting with an initial value of $\tau=0.03$.}
\end{figure}

In Fig.~\ref{fig:size_scaling_L}, we extend our analysis to larger system sizes, specifically for $L=8$ and $L=10$. We focus on an architecture with the same hyperparameters for the different sizes: number of heads ($h=10$), embedding dimension ($d=60$), linear patch size ($b=2$) and number of layers ($n_l=4$). The left panel displays the relative error of the variational energy as a function of the system size $L$ at $J_2/J_1=0.5$. The reference energies used to compute the accuracy are obtained through exact diagonalization for $L=6$~\cite{schulz1996} and through variance extrapolation from Ref.~\cite{BeccaGutz2013} and Ref.~\cite{chen2023}, for $L=8$ and $L=10$, respectively. This plot demonstrates that the accuracy remains size-consistent across the tested clusters, showing a constant behavior when increasing the system size, despite the fact that the network has fixed complexity. In the right panel, we present the computational time per optimization step as a function of the system size measured on a single GPU A100. The data illustrate how the efficiency gap between the Factored attention mechanism and the other attention mechanisms becomes more pronounced when increasing the system size.
\begin{table}[b]
    \centering
    \begin{minipage}{0.45\textwidth}
        \centering
        \begin{tabular}{cccc}
            \toprule
            & Energy & Parameters & Time  \\
            \midrule
            T5        & -0.503182(9) & 184,260    & 10h   \\ 
            Decoupled & -0.503243(9) & 184,260    & 10h    \\
            Factored  & -0.503216(8) & 154,980    & 6h   \\
            \bottomrule
        \end{tabular}
    \end{minipage}%
    \hfill
    \begin{minipage}{0.45\textwidth}
        \centering
        \begin{tabular}{cccc}
            \toprule
            & Energy & Parameters & Time  \\
            \midrule
            T5        & -0.497025(6) & 184,900    & 28h   \\ 
            Decoupled & -0.497108(6) & 184,900    & 28h    \\
            Factored  & -0.497184(6) & 155,620    & 12.5h   \\
            \bottomrule
        \end{tabular}
    \end{minipage}
    \caption{\label{table:energies} Results for the ${J_2}$-${J_1}$ Heisenberg model at $J_2/J_1=0.5$ obtained using a ViT architecture with a number of heads $h=10$, embedding dimension $d=60$, linear patch size $b=2$ and a number of layers $n_l=4$ on a $6\times 6$ (left) and on a $10\times 10$ lattice (right).}
\end{table}

In Table~\ref{table:energies} we report the results on a \(6\times 6\) and a $10 \times 10$ lattice at ${J_2/J_1=0.5}$, obtained using a four-layer architecture. In both tables, the first column shows the final mean energy achieved by the different attention mechanisms, the second column indicates the number of parameters employed in the architectures, and the last column presents the total computational time measured on a single GPU A100 to perform $5\times 10^3$ optimization steps.
It is worth noting that the accuracy of the results can be further enhanced by restoring the physical symmetries of the model through quantum number projection approaches~\cite{nomura2021,reh2023}; however, this goes beyond the scope of our work.

\begin{figure}[t]
    \center
     \includegraphics[width=\columnwidth]{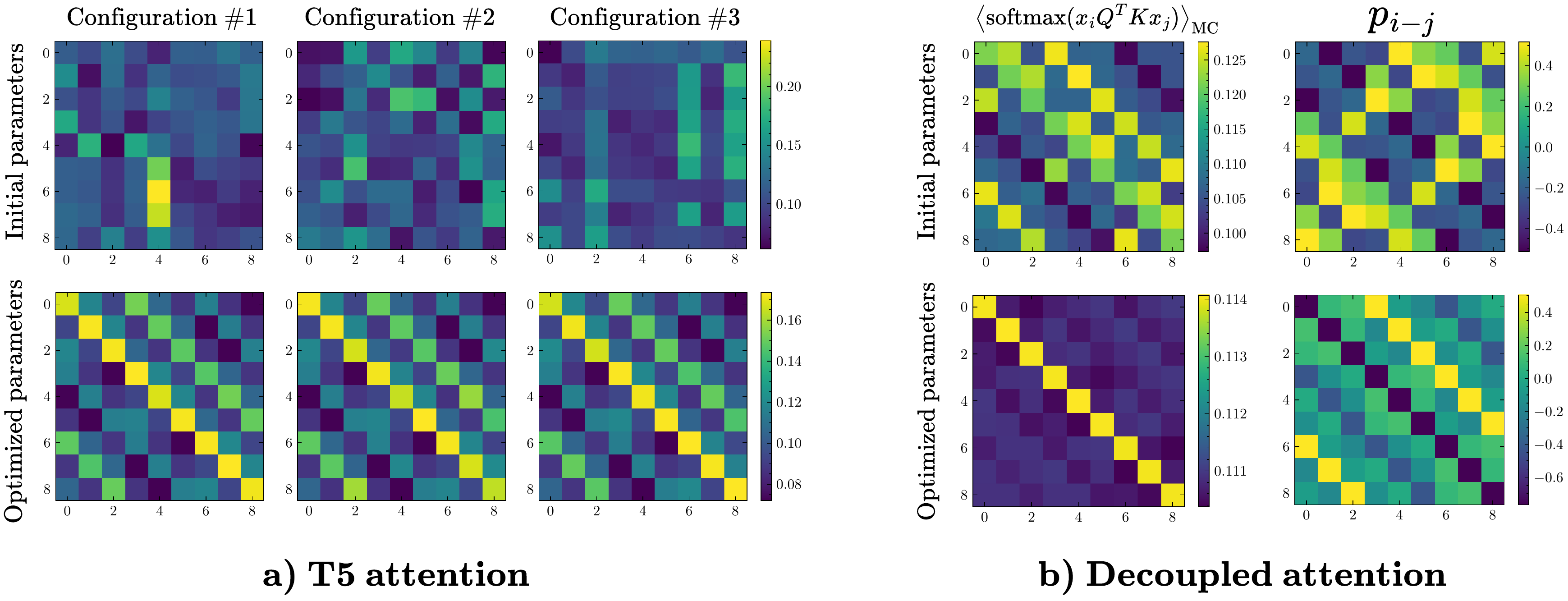}
     \caption{\textbf{Panel a:} Visualizations of the attention maps of a ViT with T5 attention mechanism [see Eq.~\eqref{eq:t5}] for three different input spin configurations. When using initial random parameters there is a clear input dependence in the attention maps (top row). Instead, at the end of the optimization, the attention maps are practically input independent (bottom row). \textbf{Panel b:} Visualizations of the input-dependent term (left panels) and of the input-independent term (right panels) of a ViT with Decoupled attention mechanism [see Eq.~\eqref{eq:decoupled}]. After the optimization (bottom row), the input-dependent term is approximately the identity matrix shifted element-wise by a constant, thus Factored attention is recovered [see Eq.~\eqref{eq:factored}]. In the plots, the input-dependent term has been averaged over $M=6\times10^3$ input configurations sampled from the optimized state. The presented results are obtained by optimizing a ViT architecture with a single layer $n_l=1$, embedding dimension $d=60$ and $h=10$ different heads on a $6\times 6$ lattice at $J_2/J_1=0.5$ (see panel (c) of Fig.~\ref{fig:opt}). The linear patch size is taken to be $b=2$, thus we have $n=9$ patches and the resulting attention maps have shape $9\times 9$. The plots are obtained by averaging the attention weights over all heads.} \label{fig:attn_maps}
\end{figure}
\subsection{Analysis of the attention maps}
The main result of the numerical simulations reported in Figs.~\ref{fig:opt},~\ref{fig:size_scaling_L} and discussed in section~\ref{subsec:numerical_experiments} is that, using a ViT employing T5, Decoupled and Factored attention, the final accuracy is practically the same (see Table~\ref{table:energies}). This suggests that, in the case of T5 and Decoupled attention, queries and keys are effectively not used in the optimized solution. To validate this statement, we study the attention maps. For the analysis, we used a single-layer architecture, where the interpretation of the results is simplified since the patches are only mixed within the attention mechanism, and the subsequent MLP cannot modify the relative weights among the various attention vectors. In panel $(a)$ of Fig.~\ref{fig:attn_maps}, we consider the case of T5 attention, plotting the attention weights defined in Eq.~\eqref{eq:t5} for three different input spin configurations. We first check that at the beginning, with random parameters, the attention maps depend on the inputs (top row), ensuring that we have an unbiased initialization. In the bottom row, we show that the architecture after optimization produces input-independent attention maps, thus automatically recovering a positional-only solution. In panel $(b)$ of Fig.~\ref{fig:attn_maps}, we consider the case of Decoupled attention, plotting separately the input dependent and the positional contributions of the attention weights [see Eq.~\eqref{eq:decoupled}]. Again, after optimization the network swaps from an unbiased solution (top row) to a positional only solution (bottom row), where the input-dependent term converges approximately to the identity matrix shifted element-wise by a constant. In other words, Factored attention is spontaneously recovered from the Decoupled version (see section~\ref{subsec:attn_mechanisms}).

\subsection{Representation of physical ground states with Factored attention}\label{sec:exact_mapping}
In this section, we provide analytic calculations about the efficacy of input-independent attention mechanisms for approximating quantum states. We first examine an analytically solvable quantum many-body Hamiltonian, developing an exact mapping between its ground state and a single layer of two-headed Factored attention. Building upon this result, we extend our analysis to scenarios where the ground state lacks analytical solutions, providing insights into why attention mechanisms including queries and keys [as in Eq.~\eqref{eq:t5} and Eq.~\eqref{eq:decoupled}] should converge to positional-only solutions when studying large systems.
\begin{figure}
\centering
\includegraphics[width=0.35\textwidth]{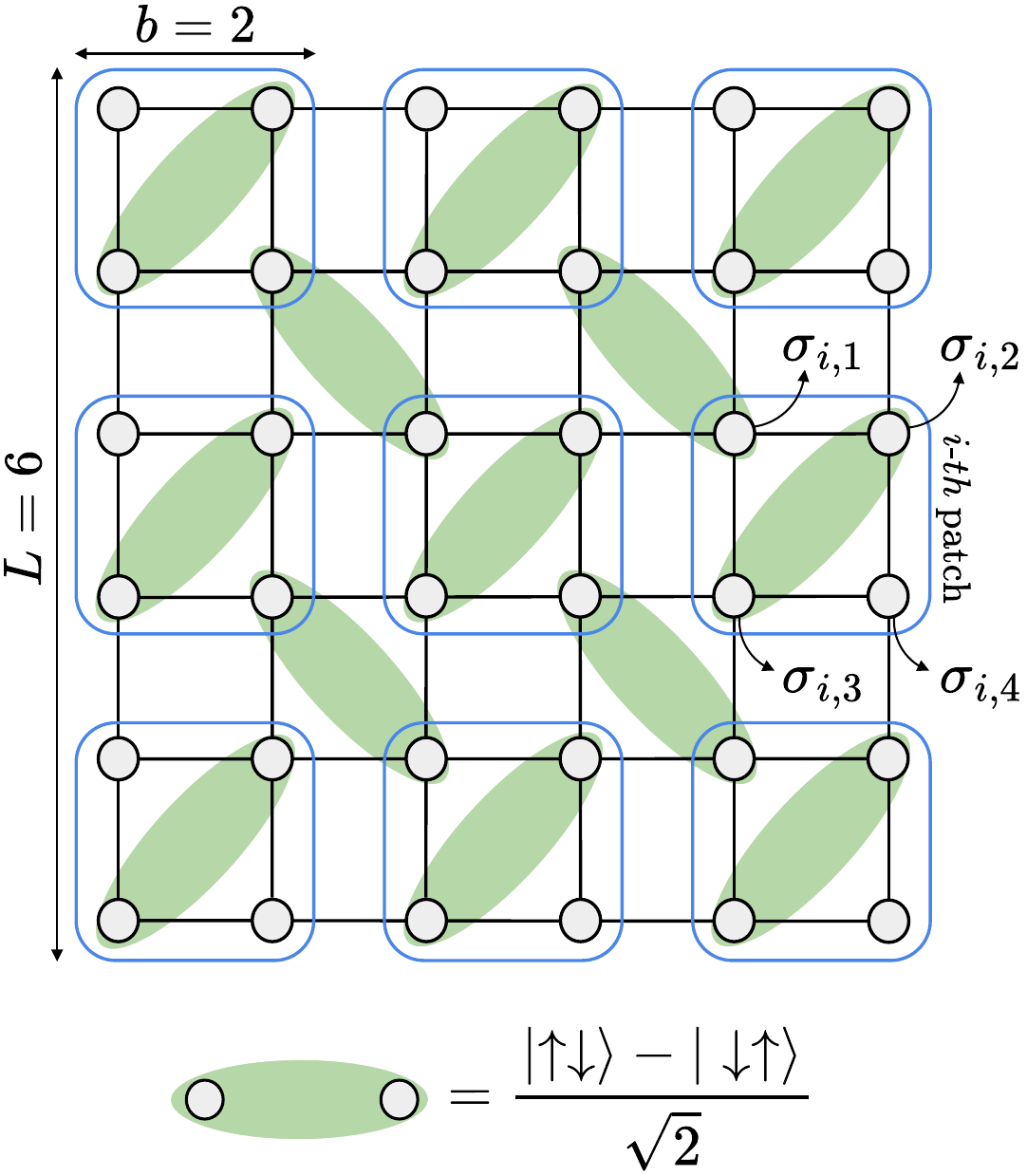}
\caption{\label{fig:shastry}Graphical representation of the ground state of the Shastry-Sutherland model in the dimer phase~\cite{shastry1981} on a \(6 \times 6\) lattice (periodic boundary connections not shown for clarity). The green shaded regions denote singlet states between two next-nearest neighbors spins. The blue squares $b \times b$ indicate the patches used to construct the input set of vectors for the Transformer.}
\end{figure}
As an illustrative example of a solvable quantum many-body Hamiltonian, we consider the Shastry-Sutherland model~\cite{shastry1981}, which captures the low-temperature properties of \(\text{SrCu}_2(\text{BO}_3)_2\), a compound known for its intriguing physical properties~\cite{Zayed_2017}. In a finite range of the frustration ratio, the ground state of this model is represented as a product of singlets between next-nearest-neighbor spins arranged on a square lattice~\cite{shastry1981}, refer to Fig.~\ref{fig:shastry} for a graphical representation. Here, we want to show that a single-layer ViT with Factored attention [see Eq.~\eqref{eq:factored}] can represent exactly this ground state. Working on a $L\times L$ square lattice with periodic boundary conditions, we partition input spin configurations into \(b \times b\) patches, with \(b = 2\) (see Fig.~\ref{fig:shastry}), which are then flattened to construct input sequences. Assuming an embedding dimension of \(d = b^2 = 4\) and choosing the embedding matrix to be the identity, the \(i\)-th input vector is \(\boldsymbol{x}_{i} = (\sigma_{i,1}, \sigma_{i,2}, \sigma_{i,3}, \sigma_{i,4})^T\), where \(i = 1, \dots, n\), with $n=L^2/b^2$.
Then, we apply the Multi-Head attention mechanism~\cite{vaswani2017} with $h=2$ heads. Considering the value matrices:
\begin{equation}
    \begin{aligned}
    V^{(1)}  = \left(\begin{array}{cccc}
0 & 0 & 0 & V_{11}^{(1)}\\
0 & V_{22}^{(1)} & V_{23}^{(1)} & 0
\end{array}\right) \  \ \ \
V^{(2)}  = \left(\begin{array}{cccc}
V_{14}^{(2)} & 0 & 0 & 0\\
0 & 0 & 0 & 0
\end{array}\right) \ ,
    \end{aligned}
\end{equation}
the value vectors are computed as $\boldsymbol{v}^{(\mu)}_{i}=V^{(\mu)}\boldsymbol{x}_{i} \in \mathbb{R}^{d/h}$:
\begin{equation}
    \boldsymbol{v}_{i}^{(1)} = \left(V_{11}^{(1)}\sigma_{i,4},V_{22}^{(1)}\sigma_{i,2}+V_{23}^{(1)}\sigma_{i,3}\right)^T \  \ \
\boldsymbol{v}_{i}^{(2)} = \left(V_{14}^{(2)}\sigma_{i,1},0\right)^T \ .
\end{equation}
Now, we assume the $n \times n$ attention matrices to be $\alpha^{(1)}_{ij}=\delta_{i,j}$ and $\alpha^{(2)}_{ij}=\delta_{i,S(i)}$, where:
\begin{equation}
    S(i) = 
    \begin{cases}
    (i + 1) \% n & \text{if} \ \ i \% (L/b) = 0, \\
    (i + L/b) \% n + 1  & \text{otherwise,}
    \end{cases}
\end{equation}
to take into account the periodic boundary conditions. Notably, the role of the two different heads is to encode the intra-patches correlations through the attention matrix $\alpha^{(1)}$ and the inter-patches correlations through $\alpha^{(2)}$. It is worth noting that, to reproduce the same attention maps with T5 [see Eq.~\eqref{eq:t5}] or Decoupled [see Eq.~\eqref{eq:decoupled}] attention mechanisms, we have to set $Q=K=0$. The resulting attention vectors are:
\begin{equation}
\boldsymbol{A}_{i}^{(1)} = \left(V_{11}^{(1)}\sigma_{i,4},V_{22}^{(1)}\sigma_{i,2}+V_{23}^{(1)}\sigma_{i,3}\right)^T \ \ \
\boldsymbol{A}_{i}^{(2)} = \left(V_{14}^{(2)}\sigma_{S(i),1},0\right)^T \ .
\end{equation}
Following the Multi-Head mechanism~\cite{vaswani2017}, we concatenate the vectors $\boldsymbol{A}_{i}^{(\mu)}$ of the different heads and apply another matrix $W \in \mathbb{R}^{d \times d}$ to mix the different representations. Choosing $W$ to be:
\begin{equation}
    W=\left(\begin{array}{cccc}
1 & 0 & 1 & 0\\
0 & 1 & 0 & 0\\
0 & 0 & 0 & 0\\
0 & 0 & 0 & 0
\end{array}\right) \ ,
\end{equation}
we obtain:
\begin{equation}
    \boldsymbol{A}_{i} =\left(V_{11}^{(1)}\sigma_{i,4}+V_{14}^{(2)}\sigma_{S(i),1},V_{22}^{(1)}\sigma_{i,2}+V_{23}^{(1)}\sigma_{i,3},0,0\right)^T \ .
\end{equation}
At this point, in the standard architecture each attention vector is fed to a MLP; in our analytical computations, we substitute it with a generic nonlinearity \(F(\boldsymbol{A}_i + c)\), where \(c\) is a constant bias. The output of this operation is the sequence of vectors:
\begin{equation}
\boldsymbol{y}_{i}=\left(F(V_{11}^{(1)}\sigma_{i,4}+V_{14}^{(2)}\sigma_{S(i),1}+c),F(V_{22}^{(1)}\sigma_{i,2}+V_{23}^{(1)}\sigma_{i,3}+c),0,0\right)^T \ .
\end{equation}
The hidden representation is obtained by summing all the output vectors $\boldsymbol{z}=\sum_{i=1}^{n}\boldsymbol{y}_{i}$, where $\boldsymbol{z} \in \mathbb{R}^d$:
\begin{equation}
    \boldsymbol{z} =\left(\sum_{i=1}^{n}F(V_{11}^{(1)}\sigma_{i,4}+V_{14}^{(2)}\sigma_{S(i),1}+c),\sum_{i=1}^{n}F(V_{22}^{(1)}\sigma_{i,2}+V_{23}^{(1)}\sigma_{i,3}+c),0,0\right)^T .
\end{equation}
Replacing the fully-connected network that acts on $\boldsymbol{z}$~\cite{rende2023,viteritti2023,rende2024finetuning} with a simpler sum, we get the amplitude of the input spin configuration: 
\begin{equation}
    \text{Log}[\Psi_{\theta}(\sigma)] = \sum_{i=1}^{n} \left[F(V_{11}^{(1)}\sigma_{i,4}+V_{14}^{(2)}\sigma_{S(i),1}+c) + F(V_{22}^{(1)}\sigma_{i,2}+V_{23}^{(1)}\sigma_{i,3}+c)  \right] \ .
\end{equation}
At the end, by choosing $F(\cdot)=\text{logcos}(\cdot)$ and setting \textcolor{black}{$V_{11}^{(1)}=V_{23}^{(1)}=\pi/4$}, $V_{14}^{(2)}=V_{22}^{(1)}=3\pi/4$ and $c=\pi/2$ we obtain an exact representation that fully complies with the ground state of the model, specifically a product of singlets arranged on a square lattice, as illustrated in Fig.~\ref{fig:shastry}:
\begin{equation}  \Psi_{0}(\sigma)=\prod_{i=1}^{L^2/4}\cos\left(\frac{\pi}{2} + \pi(\sigma_{i,4} + 3\sigma_{S(i),1})\right) \cos\left(\frac{\pi}{2} + \pi(\sigma_{i,2} + 3\sigma_{i,3})\right) \ .
\end{equation}
We want to emphasize that, to keep the analytical calculation manageable, we did not to include Layer Norm and skip connections. The mapping between the exact ground state of the Shastry-Sutherland model and the Transformer wave function highlights the role played by the different components of the architecture. In particular, this example reveals that the attention weights are used to describe the correlations in the ground state, and the attention weights connecting two patches containing uncorrelated spins should be zero to have an exact representation of the ground state. 

In general, physical events that are sufficiently far apart (either in space or time) are essentially independent or uncorrelated. From a mathematical perspective, this fundamental concept is formalized through the \textit{cluster property}~\cite{wichmann1963,Weinberg1999}:
\begin{equation}\label{eq:cluster_p}
    \lim_{|i-j|\to +\infty} \braket{\hat{B}_i\hat{B}_j} =\braket{\hat{B}_i}\braket{\hat{B}_j} \ ,
\end{equation}
where $\hat{B}_i$ is a generic local operator. According to the cluster property, correlations must decay with distance and, in the thermodynamic limit, sites that are infinitely distant become uncorrelated.
As shown in the previous mapping, the role of the attention weights is to connect correlated inputs. Therefore, for systems for which the property in Eq.~\eqref{eq:cluster_p} holds, we expect the attention weights connecting spins far apart in the system to be close to zero, regardless of the specific values of the spins. Interestingly, to reproduce this long-distance behavior using standard T5 [Eq.~\eqref{eq:t5}] or Decoupled [see Eq.~\eqref{eq:decoupled}] attention mechanisms we have to require $Q=K=0$. In other words, the standard attention mechanisms should converge to positional only solutions, thereby to the Factored version [see Eq.~\eqref{eq:factored}]. This argument, which exploits only the correlations among the elements of the input sequence, can be extended to any domain provided that the input sequence is long enough that correlations decay significantly within the scale of the system. For example, even in NLP or in computer vision tasks, when considering long input sentences or large images, it must be true that words or patches of pixels that are really far apart are uncorrelated, and so in this limit queries and keys should be optimized to zero. However, when dealing with finite sequences, this argument can have a marginal impact, and using input dependent attention weights as in Eq.~\eqref{eq:t5} can provide a good inductive bias for solving the task.

\section{Conclusion}
In this work, we showed that, when training a Transformer to approximate ground states of quantum many-body Hamiltonians, the standard attention mechanism yields equivalent performance to a simplified version, the Factored attention. The latter utilizes input-independent attention weights, resulting in fewer parameters and reduced computational cost. Moreover, starting from analytical computations, we established a direct link between attention weights and correlations. We observed that if the dominant correlation lengths necessary to solve a specific task are shorter than the total input size, the weights in conventional attention mechanisms (e.g., T5~\cite{raffel2023exploring}) should converge towards input-independent solutions. 
Interestingly, the same considerations can be extended to NLP and computer vision domains. For example, in image classification tasks, the pertinent scale is associated with the extension of objects requiring detection, typically smaller than the entire image. A straightforward approach to mitigate potential problems associated with the relationship between long-range behavior of correlations and queries and keys is the implementation of \emph{local} attention mechanisms, wherein attention weights beyond a specified distance are manually set to zero. In Ref.~\cite{rae2020transformers} it has been found that it is possible to use short-range attention for the majority of layers in the Transformer and recover the same performance of long-range language modeling. However, we emphasize that a necessary condition for the validity of our results is the possibility to partition the input sequences into effectively uncorrelated segments. This requirement may not hold universally across NLP applications. For instance, studies have demonstrated that correlations can extend over arbitrarily long scales in literary texts~\cite{altmann2012, lacalle2006}, and that, for specific tasks, global token mechanisms are preferred~\cite{qin2023nlptaskeffectivenesslongrange}. An interesting future direction of research could be the design of attention mechanisms that are able to describe the decay of long-range correlations without the necessity to set queries and keys to zero or without employing local attention mechanisms. 

\section*{Reproducibility} The variational quantum Monte Carlo and the ViT architecture were implemented in JAX~\cite{jax2018github}. The implementation of the Stochastic Reconfiguration~\cite{rende2023} is available on NetKet~\cite{netket3} under the name of \href{https://netket.readthedocs.io/en/latest/api/_generated/experimental/driver/netket.experimental.driver.VMC_SRt.html#netket.experimental.driver.VMC_SRt}{\texttt{VMC\_SRt}}. The ViT architecture used in this paper is available at \href{https://zenodo.org/records/14060431}{https://zenodo.org/records/14060431}.

\section*{Acknowledgments}
We thank A. Laio and F. Becca for useful discussions. We acknowledge the CINECA award under the ISCRA initiative, for the availability of high-performance computing resources and support.

\clearpage
\appendix
\section{Monte Carlo expectation values}\label{app:monte_carlo}
The expectation value of a quantum operator $\hat{B}$ on a variational state $\ket{\Psi_{\theta}}$ can be computed as
\begin{equation}\label{eq:exp_values}
    \braket{\hat{B}} = \frac{\braket{\Psi_{\theta}|\hat{B}|\Psi_{\theta}}}{\braket{\Psi_{\theta}|\Psi_{\theta}}} = \sum_{\{\sigma\}} P_{\theta}(\sigma) B_L(\sigma) \ ,
\end{equation}
where $P_{\theta}(\sigma) = |\Psi_{\theta}(\sigma)|^2/\braket{\Psi_{\theta}|\Psi_{\theta}}$ and $B_L(\sigma) = \braket{\sigma|\hat{B}|\Psi_{\theta}}/\braket{\sigma|\Psi_{\theta}}$ is the so-called \textit{local estimator} of $\hat{B}$. The previous expression allows us to introduce a controlled approximation method for computing expectation values. Specifically, we can perform a stochastic estimation:
\begin{equation}
     \bar{B} =\frac{1}{M}\sum_{i = 1}^{M} B_L(\sigma_i) \ ,
\end{equation}
with $\{\sigma_1, \sigma_2, \dots, \sigma_M\}$ generated from the distribution $P_{\theta}(\sigma)$ (see Appendix~\ref{app:metropolis}).
The accuracy of the estimation is controlled by a statistical error which scales as $O(1/\sqrt{M})$.

It is important to note that the computation of the local estimator $B_L(\sigma)$ in principle requires a summation over an exponential number of terms in the system size:
\begin{equation}
    B_L(\sigma) = \sum_{\{\sigma'\}} \braket{\sigma|\hat{B}|\sigma'}\frac{\Psi_{\theta}(\sigma')}{\Psi_{\theta}(\sigma)} \ .
\end{equation}
However, for local operators, such as the Hamiltonian,  $B_L(\sigma)$ can be computed efficiently. This is because the number of connected configurations $\sigma'$ for which $\braket{\sigma|\hat{B}|\sigma'}\neq 0$ scales polynomially with the system size.

\section{Metropolis Algorithm}\label{app:metropolis}
The Metropolis algorithm allows the generation of a Markov Chain~\cite{becca2017} of configurations $\{\sigma_1, \sigma_2, \dots, \sigma_M\}$ that are distributed according to $P_{\theta}(\sigma) = |\Psi_{\theta}(\sigma)|^2/\braket{\Psi_{\theta}|\Psi_{\theta}}$, without the knowledge of the normalization constant $\braket{\Psi_{\theta}|\Psi_{\theta}}$.Let us assume that $\sigma$ is the current configuration of the Markov chain. To obtain the new configuration according to the Metropolis algorithm, we perform the following steps:
\begin{enumerate}
    \item Generate a configuration $\sigma'\sim k(\sigma'|\sigma)$, where $k(\sigma'|\sigma)$ is the \textit{proposal kernel}~\cite{becca2017}.
    \item Evaluate the log-acceptance ratio of the proposed move:
    \begin{equation}
        \text{log}[A(\sigma',\sigma)] = \text{min}\left(0, \text{log}\left[\frac{P_{\theta}(\sigma')}{P_{\theta}(\sigma)}\right]\right) \ ,
    \end{equation}
    where
    \begin{equation}
        \text{log}\left[\frac{P_{\theta}(\sigma')}{P_{\theta}(\sigma)}\right] = 2\Re\{\text{Log}[\Psi_{\theta}(\sigma')]\} - 2\Re\{\text{Log}[\Psi_{\theta}(\sigma)]\} \ ,
    \end{equation}
    \item Accept the new configuration $\sigma'$ with probability $A(\sigma',\sigma)$. In practice, this is done by drawing a random number $u \in (0,1]$ and proceeding as follows:
    \begin{itemize}
        \item \textbf{Accept} the move if $\text{log}(u) \leq \text{log}[A(\sigma',\sigma)]$;
        \item \textbf{Reject} the move if $\text{log}(u) > \text{log}[A(\sigma',\sigma)]$, in this case the new configuration in the Markov Chain remains $\sigma$.
    \end{itemize}
\end{enumerate}
Notice that the described formulation of the Metropolis algorithm relies solely on the logarithm of the wave function $\text{Log}[\Psi_{\theta}(\sigma)]$. This is useful from a practical standpoint to avoid numerical issues, such as underflow and overflow, when evaluating the non-normalized wave function.

In the case of the $J_1$-$J_2$ Heisenberg model studied in this work, due to the $SU(2)$ spin symmetry of the Hamiltonian, the total magnetization is conserved and the  ground-state search can be limited in the $S^z = 0$ sector. This can be implemented in the Monte Carlo sampling by proposing the flipping of two spins oriented in opposite directions when generating the new configuration $\sigma'$ (see step (1) of the Metropolis algorithm).

\clearpage
\bibliography{refs}
\bibliographystyle{unsrt}

\end{document}